\documentclass[%
 reprint,https://www.overleaf.com/project/619cf66669f709601a2664e7
 amsmath,amssymb,
 aps,
]{revtex4-2}

\usepackage{graphicx}
\usepackage{dcolumn}
\usepackage{bm}
\usepackage{xcolor}
\usepackage{version}
\usepackage{tikz}
\usetikzlibrary{shapes.geometric}
\usepackage{amsmath}
\usepackage{appendix}

\begin{document}

\preprint{APS/123-QED}

\title{Fixed Point Quantum Monte Carlo}

\author{Romain Chessex}
\email{rchessex@ethz.ch}
\author{Massimo Borrelli}%
\author{Hans Christian \"{O}ttinger}

\affiliation{Polymer Physics, Department of Materials, ETH Z\"{u}rich, CH-8093 Z\"{u}rich, Switzerland}%




\date{\today}

\begin{abstract}
We present a new approach to the study of equilibrium properties in many-body quantum physics. Our method takes inspiration from Density Matrix Quantum Monte Carlo and incorporates new crucial features. First of all, the dynamics is transferred to the Laplace representation where an exact equation can be derived and solved using a simulation-step that, unlike most Monte Carlo methods, is not \textit{a priori} physically bounded. Moreover, the spawning events are formulated in terms of two-process stochastic unravellings of quantum master equations, a formalism that is particularly useful when working with density matrices. And last, this is equivalent to an interaction picture, where the free part is integrated exactly and the convergence rate can be greatly increased if the interaction parameter is small. We benchmark our method by applying it to two case-studies in condensed matter physics, show its accuracy and further discuss its efficiency.
\end{abstract}

\maketitle

\section{Introduction}
Quantum Monte Carlo (QMC) methods have been established as a leading paradigm for numerical simulations in computational quantum physics, with successful applications in condensed matter as well as in quantum chemistry \cite{BeccaSorella,RevQMC}. Different approaches have been developed whose applicability can depend strongly on the specific physical or chemical system under investigation.
 Variational Quantum Monte Carlo (VQMC) \cite{teller,assaraf} and Projector Quantum Monte Carlo (PQMC) \cite{pqmc, shift_update} are probably the most common methods for studying the zero-temperature properties of highly-correlated many-body systems, though they greatly differ from one another. While in VQMC the ground state energy of is obtained via a minimization procedure starting from an initial guess of the ground state function, PQMC methods, such as Diffusion Monte Carlo and Green Function Monte Carlo  \cite{ulam,kalos,dqmc,gfqmc}, rely on an iterative stochastic projection that will drive the state of the system to its true ground state. 
 For non-zero temperature systems, Auxiliary Field Monte Carlo (AFMC) \cite{SUGIYAMA19861,lang,Saidi} and Path Integral Monte Carlo (PIMC) \cite{feynman} have been formulated, based on thermal field theory, to investigate lattice models and quantum statistical mechanics, respectively. While for bosons QMC methods have proven extremely successful and have provided nearly exact results \cite{boson1,boson2,boson3}, simulations for fermionic systems have long been hampered by the well-known sign-problem \cite{sse, h3,sgn_prob1, sgn_prob2}, a numerical artifact arising as a direct consequence of the anti-symmetric properties of the wave-function. The fixed-node approximation \cite{fixednode,fixednode2,fixednode3} based on guessing the nodal structure for an initial trial function, has given some good results, although its applicability is strongly limited by the feasibility to effectively assess such a structure prior to the simulation.\par
In \cite{FCIQMC1,FCIQMC2} a new projective QMC method was introduced to investigate the zero-temperature properties of correlated electrons. This method, named Full Configuration Interaction Quantum Monte Carlo (FCIQMC), relies on simulating stochastic trajectories in the space of Slater determinants, thus not requiring any prior knowledge about the nodal structure of the many-body wave-function and allowing one to tackle the sign problem more directly, based on efficient annihilation of undesired Monte Carlo walkers \cite{FCIQMC3, nature_fciqmc}. A few years later, a density matrix method (DMQMC) inspired by FCIQMC was introduced in \cite{DMQMC} and applied to study finite temperature behaviour of correlated fermions \cite{dmqmc_ueg, ipdmqmc}. Here, the authors devised a Monte Carlo algorithm that simulates trajectories in the space of operators, rather than quantum states, and it therefore offers a natural recipe to sample stochastically the relevant density matrix elements independently. Moreover, using this method allows for a direct evaluation of quantities such as entanglement and correlations in general. Recently, this approach was extended to open quantum systems described by a Lindblad-type master equation and applied to dissipative quantum magnetism \cite{nagy}.\par Here, we adopt a DMQMC approach and combine it with a field theory framework to study equilibrium properties of strongly interacting many-body systems. Starting from the general theoretical framework originally introduced in \cite{hansfieldtheory} and later fully illustrated in \cite{PhilApp}, we design a Quantum Monte Carlo algorithm based on two-process stochastic unravellings \cite{breuer, PhilApp} to solve the symmetrized Bloch equation. The main advantage of our approach, as opposed the standard DMQMC, lies in the use of a completely general and exact methodology describing the dynamics of the fully interacting many-body system in the Laplace space. Most importantly, our framework can be considered as an interaction picture, in the sense that the system evolves in the basis of the free Hamiltonian. This is obviously of tremendous help since  in free field theories the eigenbasis is virtually always known {\itshape a priori} and its evolution can be integrated exactly. Moreover, this provides a recipe for numerical integration in which no physical bounds on the simulation time-step are present. This is in stark contrast to standard projective methods where the total width of the many-body spectrum usually sets a fundamental limit. From an implementation perspective, our algorithm includes a series of standard MC features, such as importance sampling and approximations, that decrease the statistical errors and speed-up the simulations. To demonstrate the efficiency and range of applicability of our method we study the ground-state convergence in two well-known models in condensed matter physics, namely the two-dimensional Heisenberg XXZ model \cite{heisenberg} and the Fermi-Hubbard model \cite{hubbard}.
\par
This article is organised as follows. In Sec.~\ref{Theory} we illustrate the theoretical basis of our algorithm. In 
Sec.~\ref{Algo} we introduce the QMC algorithm itself and explain the main steps in detail. In Sec.~\ref{sec:results} we test the algorithm and discuss our findings. Finally, in Sec.~\ref{Concs} we draw some conclusions and outline some open questions for future investigations.

\section{Theoretical foundations of the algorithm}
\label{Theory}
In this section we illustrate the main theoretical ideas behind our Monte-Carlo approach. First, we will illustrate the fundamental equation describing the evolution of a quantum system toward its ground state. After this, we shall introduce the stochastic unravelling approach that will serve as the starting point for the algorithmic implementation.
\subsection{The deterministic model}
The typical scenario we want to address is a that of a many-body quantum system whose dynamics is dictated by 
the Hamiltonian
\begin{equation}
H=H^{\textrm{free}}+H^{\textrm{int}},
\label{Ham1}
\end{equation}
where the eigensystem of the free Hamiltonian $H^{\textrm{free}}$ is fully known and $[H^{\textrm{free}},H^{\textrm{int}}]\ne0$. Since we want to work with the density matrix formalism, all the following discussions will be entirely formulated in terms of superoperators to describe the time-evolution of quantum states. Generally speaking, the time evolution of the density matrix $\rho$, describing the state of the many-body system, is dictated by Von Neumann equation
\begin{equation}
\frac{d}{dt}\rho=-i[H,\rho].
\label{eq:vonNeumann}
\end{equation}
As we are interested in ground state properties, similarly to previous projective methods \cite{FCIQMC1,DMQMC} we consider the following imaginary-time symmetric equation, known as Bloch equation \cite{DMQMC} 
\begin{equation}
\frac{d}{d\beta}\rho=-\frac{1}{2}\{H,\rho\}=\mathcal{L}\rho,
\label{bloch}
\end{equation}
with $\beta=it$. 
If one introduces an energy shift
\begin{equation}
    H \mapsto H - S_\rho, \qquad S_\rho = \frac{\text{Tr}(H \rho)}{\text{Tr}(\rho)},
    \label{eq:bloch_shift}
\end{equation}
Eq.\ (\ref{bloch}) becomes a legitimate master equation, with trace-preserving properties. Adding this energy shift results in an extra term $S_{\rho}\rho$ on the right and side of Eq.\ (\ref{bloch}). In doing that, one trades the advantage of a trace-preserving equation at the expense of introducing non-linearity in the original equation. The general solution of the shifted Bloch equation \eqref{eq:bloch_shift} reads
\begin{equation}
\rho_{\beta}=e^{-\frac{\beta}{2}(H-S_{\rho})}\rho_{0}e^{-\frac{\beta}{2}(H-S_{\rho})}.
\label{gensolbloch}
\end{equation}
If the ground-state of the system is non-degenerate, when $\beta\to\infty$, its contribution to the above expansion will become the dominant one, that is
\begin{equation}
 \lim_{\beta\to\infty}\rho_{\beta}\propto|E_{0}\rangle\langle E_{0}|,
 \label{groundstateconv}
\end{equation}
which implies that $S_{\rho}\to E_{0}$ consistently.
The non-linearity in Eq.~\eqref{eq:bloch_shift} makes the solution of the Bloch equation a non-trivial problem that might require further approximations when one is not just interested in steady state properties.
For that reason, we do not solve the evolution equations in full, but rather rely on a modified version of the stationary condition. Starting from the observation that the most general solution Eq.~ (\ref{gensolbloch}) can be re-expressed in terms of the one-sided dynamical generator $\mathcal{L}$, that is
\begin{equation}
\rho_{\beta}=e^{\mathcal{L}\beta}\rho_{0},
\label{gensolblochL}
\end{equation}
the following stationary condition can be easily derived 
\begin{equation}
{\mathcal{L}}\rho_{\infty}=0,
\label{groundstateconvL}
\end{equation}
by imposing $d{\rho}/d\beta=0$, with $\rho_{\infty}$ being the steady state. If we multiply both side of Eq.~(\ref{groundstateconvL}) by an inverse temperature scale $1/r$ we can recast this equation in the following dimensionless form  
\begin{equation}
\left(1+\frac{\mathcal{L}}{r}\right)\rho_{\infty}=\rho_{\infty},
\label{eq:groundstateconvL2}
\end{equation}
and by decomposing the total Liouvillian superoperator $\mathcal{L}$ into its free and interaction part, $\mathcal{L}^{\text{free}}$ and $\mathcal{L}^{\text{int}}$ respectively, Eq.~(\ref{eq:groundstateconvL2}) can be further rewritten as
\begin{equation}
    r\left(1 + \frac{\mathcal{L}^{\text{int}}}{r}\right)\rho_\infty = \left(r-\mathcal{L}^{\text{free}}\right)\rho_\infty.
\label{eq:groundstateconvL3}
\end{equation}
The right hand side of the above equation is the inverse of the Laplace transform of $e^{\beta\mathcal{L}^{\text{free}}}$, i.e.
\begin{equation}
    \mathcal{R}_r^{\text{free}} = \int_0^{\infty} e^{\beta\mathcal{L}^{\text{free}}}e^{-r\beta}d\beta = \frac{1}{r-\mathcal{L}^{\text{free}}}.
\end{equation}
Once replaced in Eq.~(\ref{eq:groundstateconvL3}) it leads to 
\begin{equation}
    r\mathcal{R}_r^{\text{free}}\left(1+\frac{\mathcal{L}^{\text{int}}}{r}\right)\rho_\infty = \rho_\infty,
    \label{eq:fund}
\end{equation}
which is the fundamental equation for all our ground state calculations.

\subsection{Two-process stochastic unravelling}
All the results illustrated so far are exact and represent a continuous and deterministic description of the density matrix evolution. Needless to say, for multi-particle systems with intricate interactions, a numerically exact solution of Eq.~(\ref{eq:fund}) is practically out of reach, owning to the huge dimension of the Hilbert space. However, by a successive application of the Eq.~(\ref{eq:fund}), one gets
\begin{equation}
    \left[r\mathcal{R}_r^{\text{free}}\left(1+\frac{\mathcal{L}^{\text{int}}}{r}\right)\right]^{n}\rho_\infty = \rho_\infty,
    \label{eq:fund_disc}
\end{equation}
where $r$ is fixed, setting the inverse temperature resolution. Obviously, in the zero temperature limit, the ground-state will emerge as the solution to Eq.~(\ref{eq:fund}) and its iterated version (\ref{eq:fund_disc}) as well. Starting from this observation a stochastic unravelling, which we named \textit{triplet unravelling}, can be developed. This relies on stochastic trajectories in the Hilbert space that are represented by triplets of the form $(c,|\phi\rangle, |\psi\rangle)$ 
with $c$ being a complex number.  The piece-wise-deterministic stochastic processes will alternate between continuous, exact free evolution, as dictated by $H^{\textrm{free}}$,  interrupted by random quantum jumps (or collisions) associated to $H^{\textrm{int}}$. In order for this method to provide a statistically robust solution to Eq.~\eqref{eq:fund_disc}, the following equation must hold
\begin{equation}
    \rho_{\infty} = \mathbb{E}[c|\phi\rangle\langle \psi|],
\end{equation}
in which $\mathbb{E}$ represents a statistical average over all the trajectories. In other words, the solution to the exact Eq.~\eqref{eq:fund}, \textit{i.e.} the ground state, must be recovered. Since we are working with the density matrix formalism and we aim at generating trajectories that are the least possible statistically correlated, we model two-side collision processes as follows 
\begin{equation}
|\phi\rangle\langle \psi|\rightarrow|\phi\rangle\langle \psi|-\frac{1}{2r}\left(H^{\textrm{int}}|\phi\rangle\langle \psi|+|\phi\rangle\langle \psi|H^{\textrm{int}}\right),
    \label{eq:collisions}
\end{equation}
which can be interpreted as a stochastic implementation of the operator $\left(1+\frac{\mathcal{L}^{\text{int}}}{r}\right)$ with quantum jumps occurring between connected states at a rate $r$. As for the free evolution, we assume this can be solved exactly which is always the case if the free Hamiltonian $H^{\textrm{free}}$ is fully known and its eigenstates are used as basis for the triplet realizations.

\section{Algorithmic realization of triplet unravelling}
\label{Algo}
In this section we present our algorithm. For the sake of clarity and readability, the original features of our algorithm are explained thoroughly, while in the Appendices we discuss some aspects that our method shares with DMQMC and FICQMC as well as other minor technicalities. We introduce the following short-hand notation for a triplet $(c,|\phi\rangle, |\psi\rangle)\equiv(c,\phi,\psi)$, which will be adopted in all following discussions.
\subsection{The Monte-Carlo walkers: triplets}
Like any other Monte-Carlo method, ours too statistically samples the density matrix representing the steady state via an ensemble of walkers. These are chosen as an ensemble of triplets $\{(c_n,i_n, j_n)\}_n$, where $i_n, j_n$ are local basis vectors (\textit{e.g.}, the free Hamiltonian eigenstates) and the $c_{n}$ weights are, in general, complex. The evolution of this ensemble will be our Monte Carlo simulation of the piece-wise unravelling provided by Eq.~\eqref{eq:collisions} of the Bloch equation in the Laplace representation. If one generates $N$ stochastic trajectories, the density matrix is statistically reconstructed via the following average
\begin{equation}
  \rho = \frac{1}{\mathcal{N}}\sum_n c_n |i_{n}\rangle\langle j_{n}|,
    \label{eq:DMestimation}
\end{equation}
where $n$ labels a single trajectory and the normalization $\mathcal{N} = \sum_n c_n \text{Tr}(|i_{n}\rangle\langle j_{n}|)$ ensures that density matrix has the correct trace. Note that the addition of the normalization solves the non-conserving trace problem of the shifted Bloch equation. After initializing the density matrix ensemble to the free Hamiltonian ground state, the algorithm develops in a series of identical loops, each loop consisting of two main steps, spawning events, realized via quantum jumps, and continuous free evolution. As we shall show, the use of a discrete basis, combined with a signed weight for the triplets, will be the key for an effective cancellation of positive and negative contributions to averages, allowing to reduce the sign problem. Finally, statistical quantum averages of operators can be easily calculated using the ensemble statistics. Using Eq.~\eqref{eq:DMestimation} one finds that for a general  operator $A$ and an ensemble $\{(c_n,i_n,j_n)\}_n$ the quantity $\text{Tr}(A\rho)$ at the end of each loop iteration can be estimated as
\begin{equation}
    \text{Tr}(A\rho) = \frac{\sum_n c_n a_{i_nj_n}}{\sum_n c_n \delta_{i_nj_n}},
    \label{eq:measure}
\end{equation}
where $a_{i_nj_n} = \text{Tr}(A|i_{n}\rangle\langle j_{n}|)$. In order to illustrate better the ensemble normalisation, we define the \textit{population of the ensemble} as the sum of the absolute weight over all the triplets. 
\subsection{The main loop}
The algorithm starts by initializing the statistical ensemble to the free Hamiltonian ground state, that is 
$(c_{\text{init}}, |e_0\rangle, \langle e_0|)$, with $c_{\text{init}} > 0$. 
If the free ground state is degenerate, $N_{\text{i}}\leq N_{\text{GS}}^f$ triplets are chosen uniformly among the $N_{\text{GS}}^{f}$ possible ground states. In what follows we illustrate the main loop of the algorithm, including the compression/decompression steps (see Appendix \ref{sec:comp}).
\\
    \paragraph{\underline{Spawning}}  
    \begin{enumerate}
        \item Pre-spawning decompression.  
        \item Spawning. For each unit-weight triplet either one of the two states is chosen randomly (quantum mechanically, either a ket or a bra). For instance, if the ket is chosen, a new state $k$ is spawned from $i_n$ with probability $p_{i_nk}$. Then the newly spawned triplet will read
        \begin{equation}
            \left[-\textup{sign}(c_n)\frac{H^{\text{int}}_{i_nk}}{r p_{i_nk}}, k, j_n\right].
            \label{eq:jump}
        \end{equation}
        Equivalently if $j_n$ is selected.
    \end{enumerate} 
    This step numerically implements the application of the super-operator $(1 + \frac{\mathcal{L}^{\text{int}}}{r})$. The new state $k$ is usually chosen uniformly among all the possible $n_s$ spawning events such that $\langle k|H^{\text{int}}|i_n\rangle \neq 0$ and whose probability is $p_{i_nk} = 1/n_s$.\\ 
    \\
    \paragraph{\underline{Free evolution}}
    \begin{enumerate}
        \item Triplets compression. 
        \item For each $(c_n,i_n,j_n)$ a weight update for the free contribution is performed according to 
        \begin{equation}
            c_n \mapsto \frac{r}{r - S + (H_{i_ni_n}^{\text{free}} + H_{j_nj_n}^{\text{free}})/2}c_n.
        \end{equation}
    \end{enumerate}
    Complementary to the previous step, this step numerically implements the application of $r\mathcal{R}_r^{\text{free}}$, the Laplace transform of the free evolution operator. If the local states $i_n$ and $j_n$ are eigenvectors of the free Hamiltonian, which will always be the case in our simulations, this step is exact and it only modifies the statistical weights of the triplets.
    The shift can then be updated. When the population has reached the desired value, the shift is updated regularly. This will have the effect to stabilize the evolution of the population. We previously defined the shift to be the average energy $\text{Tr}(H\rho)/\text{Tr}(\rho)$ but this choice turns out to be less efficient than the prescription used in the DMQMC method (see Appendix \ref{sec:shiftupdate} for more details).
    
\subsection{Importance sampling and initiator approximation}
\begin{figure}
    \centering
\begin{tikzpicture}
    \draw[->](1.5, 1+1/3)--(3,0.5);
    \draw[->](1.5, 2)--(3,2);
    \draw[->](1.5, 2+2/3)--(3,3.5);
    \filldraw[black, fill = white] (1.25, 3.6) arc [start angle = 90, end angle = -270, x radius = 1, y radius = 1.6];
    \draw[blue](1,3)--(1,1) node[below]{$c$};
    \draw[red](1.5,3)--(1.5,1) node[below]{$w$};
    \foreach \i in {0,1,...,6}{
        \draw[blue](1.1,\i/3+1)--(0.9,\i/3+1);
    }
    \foreach \i in {0,1,...,3}{
        \draw[red](1.4,\i/1.5+1)--(1.6,\i/1.5+1);
    }
\draw(1.25,0.4) node[below] {$(c,i,j)$};
    \draw[blue](3.5,0.5-1/3)--(3.5,0.5+1/3);
    \draw[blue](3.4, 0.5-1/3)--(3.6, 0.5-1/3);
    \draw[blue](3.4, 0.5)--(3.6, 0.5);
    \draw[blue](3.4, 0.5+1/3)--(3.6, 0.5+1/3);
    \draw[red](4,0.5-1/3)--(4,0.5+1/3);
    \draw[red](3.9, 0.5-1/3)--(4.1, 0.5-1/3);
    \draw[red](3.9, 0.5+1/3)--(4.1, 0.5+1/3);
    \draw[blue](3.5,2-1/3)--(3.5,2+1/3);
    \draw[blue](3.4, 2-1/3)--(3.6, 2-1/3);
    \draw[blue](3.4, 2)--(3.6, 2);
    \draw[blue](3.4, 2+1/3)--(3.6, 2+1/3);
    \draw[red](4,2-1/3)--(4,2+1/3);
    \draw[red](3.9, 2-1/3)--(4.1, 2-1/3);
    \draw[red](3.9, 2+1/3)--(4.1, 2+1/3);
    \draw[blue](3.5,3.5-1/3)--(3.5,3.5+1/3);
    \draw[blue](3.4, 3.5-1/3)--(3.6, 3.5-1/3);
    \draw[blue](3.4, 3.5)--(3.6, 3.5);
    \draw[blue](3.4, 3.5+1/3)--(3.6, 3.5+1/3);
    \draw[red](4,3.5-1/3)--(4,3.5+1/3);
    \draw[red](3.9, 3.5-1/3)--(4.1, 3.5-1/3);
    \draw[red](3.9, 3.5+1/3)--(4.1, 3.5+1/3);
    \draw(3.75, 4) arc [start angle = 90, end angle = -270, x radius = 0.7, y radius = 0.5];
    \draw(3.75, 2.5) arc [start angle = 90, end angle = -270, x radius = 0.7, y radius = 0.5];
    \draw(3.75, 1) arc [start angle = 90, end angle = -270, x radius = 0.7, y radius = 0.5];
    \draw(4.5, 0.5) node[right]{$\left(\frac{c}{|w|}, i, j\right)$};
    \draw(4.5, 2) node[right]{$\left(\frac{c}{|w|}, i, j\right)$};
    \draw(4.5, 3.5) node[right]{$\left(\frac{c}{|w|}, i, j\right)$};
\end{tikzpicture}
    \caption{Example of a triplet split $(c,i,j)$ according to the weight factors (red) with physical weight $c$ (blue). The original triplet with $|c|= 6$ and bias $b_{ij} = 2$ is split into $\lfloor|w|\rfloor = 3$ child triplets according to the weight factor. Each child triplet will then attempt a single spawning from $(\frac{c}{|w|}, i,j)$.}
    \label{fig:impsamp}
\end{figure}
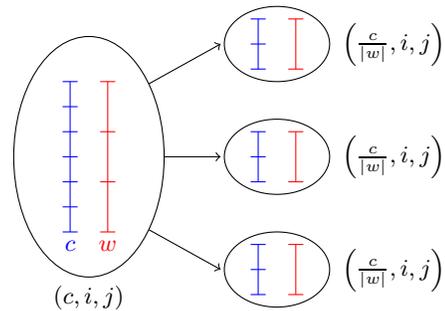

We introduce an importance sampling scheme based on a \textit{dynamic norm} $n_{ij}$, defined for a triplet $(c,i,j)$ as the minimum number of applications of the interaction Hamiltonian $H^{\textrm{int}}$ needed to jump from $i$ to $j$. This idea is rooted in the observation that, for short-range interaction systems, most of the observables average values are sampled by triplets with a short dynamic norm.
In order to reduce the variance, instead of visiting all the possible connecting states, we force spawning events towards states with a shorter dynamic norm, thus limiting the statistical exploration of the Hilbert space to relevant regions only. One can picture this as forcing the stochastic sampling to occur mostly around the main diagonal of the density matrix.
To make this idea concrete we associate to a triplet two types of weight, a \textit{physical weight} and a \textit{weight factor}. The first, denoted by $c$, is the weight that has been used until now for averages. The latter, denoted by $w$, reflects instead the number of spawning attempts that will be performed by a triplet. The two are related by a norm-dependent bias $b \equiv b(n_{ij}) > 1$ via the equation $c = bw$. In order to decrease the number of triplets with large dynamic norm the bias should increase as the norm increases. Accordingly, the decompression step is performed with respect to the weight factors $w$ as to decrease the relative number of spawning attempts associated to larger dynamic norms. This means that triplets $(c,i,j)$ are split into $\lfloor |w|\rfloor$ child triplets of weight $c/|w|$ and a rest triplet surviving with probability $|w|-\lfloor|w|\rfloor$. If the rest triplet survives, its weight is updated to $c/|w|$. Hence, those triplets associated to larger dynamic norms (corresponding to larger biases) will attempt less spawning events, keeping the simulation from explore unimportant regions of the Hilbert space. This procedure is illustrated in Fig.\ \ref{fig:impsamp}.
Up to this point, no explicit formula for the bias $b$ has been provided. Based on early discussions, it must be norm-dependent, and must increase as the $n_{ij}$ increases. We model the importance sampling as a harmonic interaction with spring constant $\kappa$ between the states $i$ and $j$ which will force them to stay dynamically close. The corresponding bias reads
\begin{equation}
    b = \exp\left(\frac{\kappa}{2}n_{ij}^2\right).
    \label{eq:spring}
\end{equation}
Note, that with that choice $b_{ii} = 1$, which implies that the initial ensemble is unbiased. 
As a side remark, we note that each spawning event requires at least the computation of two norms: one for the original state and one for the spawning. If the dynamic norm is computationally expensive, it can make the simulation very inefficient. However, another norm $\hat{n}_{ij}$ between states can still be defined in order to reduce the variance while being faster to compute and numerically close to the dynamic norm. An example of an alternate norm will be used in Sec.\ \ref{sec:results}. Note also that the population is computed with the weight factors, to reflect the correct number of spawning performed.\par
We conclude this section by illustrating our initiator approximation, which is based on the FCIQMC version in \cite{FCIQMC2}. The rationale here is to have an additional survival criterion for the newly spawned triplets that reduces the ensemble population needed for convergence. Only some triplets, labelled as \textit{initiators}, will be given the possibility to spawn other triplets that are not yet present in the original ensemble. The only exception to this rule is if two triplets spawn simultaneously the same triplet. The ensemble of initiators can increase if their weight, in absolute value, is larger than a critical value $c_{\text{init}} > 0$ or if its dynamic norm is strictly lower than a threshold $n_{\text{init}}$. 
This translates into a modification of the compression and decompression steps (see Appendix I for details).

\section{Results and discussions}
\label{sec:results}

In this section we benchmark our algorithm using two paradigmatic models in condensed matter physics; the antiferromagnetic Heisenberg model on square and triangular lattices, and the Fermi-Hubbard model on a square lattice. A typical simulation consists of repeated iterations of the main loop illustrated above and, generally speaking, it can be separated into two distinct phases: \textit{thermalization} and \textit{sampling}. We call thermalization the convergence phase from the initial state to the ground state, where the number of iterations is $N_{\text{ther}}$ and during which the fixed point solution is reached. Sampling generates instead an ensemble of stochastic fixed point solutions. The initial condition is taken from the solution of the previous one for a sequence of $N_{\text{samp}}$ estimates of the quantum average of the observable $A$, using Eq.\ (\ref{eq:measure}), is stored. Since all samples are calculated from the same trajectory at different iteration, they are correlated, i.e. the density matrices are estimated by the same statistical ensemble at different iteration. It is hence clearly necessary to take into account those correlations when calculating the statistical error.  Variance estimation techniques for correlated samples such as binning analysis \cite{binning_analysis} allow to estimate the true statistical error.

\subsection{Case study: the Heisenberg antiferromagnet}
\label{Heisenberg}
We consider a spin-$1/2$ Heisenberg model on a two-dimensional lattice \cite{heisenberg}. This is a paradigmatic model in quantum magnetism, whose exact solutions can be only be found for specific cases \cite{zvyagin} and, as such, it is still the subject of intense theoretical and numerical investigations. The general Hamiltonian reads
\begin{equation}
H = \sum_{\langle a,b\rangle} J_{x}\sigma_a^{x} \sigma_b^{x}+J_{y}\sigma_a^{y} \sigma_b^{y}+J_{z}\sigma_a^{z} \sigma_b^{z},
\end{equation}
where $\sigma^x, \sigma^y, \sigma^z$ are the standard Pauli matrices and $\langle a, b\rangle$ denotes nearest neighbours on the lattice. In our simulations, the lattice can be either triangular or squared where spins sit on the lattice points. By setting $J_{x}=J_{y}=2J_{z}=J$ the Hamiltonian of the XXZ spin model is recovered
\begin{equation}
H=\frac{J}{2}\sum_{\langle a,b\rangle}\sigma_a^+\sigma^-_b + \sigma_a^-\sigma_b^+ + \sigma_a^z \sigma_b^z,
\end{equation}
where $\sigma^\pm = \sigma^x \pm i\sigma^y$. This can be further split into a free part and an interacting part $H = H^{\textup{free}} + H^{\textup{int}}$, with
\begin{equation}
H^{\text{free}} = \frac{J}{2}\sum_{\langle a,b\rangle} \sigma_a^z \sigma_b^z, \qquad H^{\text{int}} =\frac{J}{2}\sum_{\langle a,b \rangle} \sigma_a^+\sigma^-_b + \sigma_a^-\sigma_b^+.
\end{equation}
 For a lattice of $L\times L$ spins, we introduce the eigenbasis of the free Hamiltonian
\begin{equation}
|e_{s_1,\dots,s_{L\times L}}\rangle = |s_1\rangle \otimes |s_2\rangle \otimes \dots \otimes |s_{L\times L}\rangle,
\end{equation}
where $s_a = s^z_a= \pm 1$. These describe a precise spin configuration of the lattice, where each site is either in a state with spin up or down. The single-particle operators $\sigma^z$ and $\sigma^\pm$ act of the basis states according to the standard algebra of Pauli matrices 
%
\begin{equation}
\sigma^z|\pm 1\rangle = \pm|\pm 1\rangle, \qquad \sigma^\pm  |\mp 1\rangle = |\pm 1 \rangle, \qquad \sigma^\pm |\pm 1\rangle = 0.
\end{equation}
Note that the interaction Hamiltonian does not change the total spin, which is therefore a conserved quantity. We can hence restrict our attention to subspaces of the total Hilbert space characterized by states with an equal number of up and down spins to find the ground state.
\begin{figure}
    \centering
    \includegraphics[width = \linewidth]{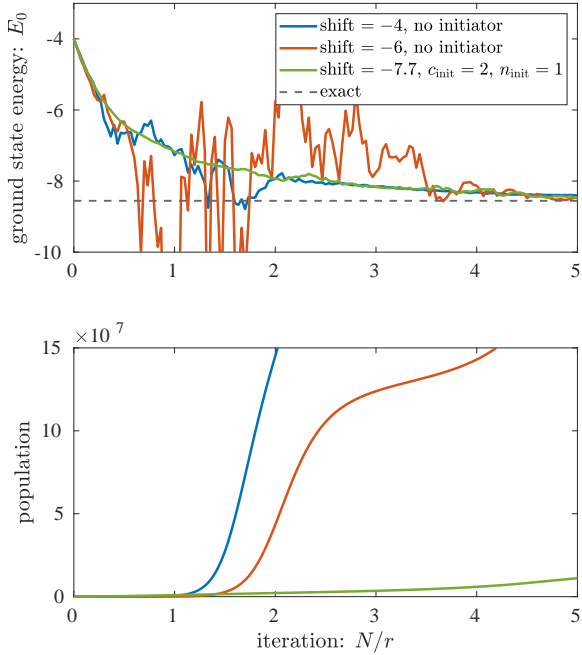}
    \caption{Ground state thermalization for the $4\times 4$ triangular Heisenberg model for $r = 30$, $N_{\text{i}} = 10^3$ and various constant shift values. The plateau stage is clearly visible. The exact ground state density matrix has approximately $1.6\cdot 10^8$ elements.}
    \label{fig:sgn_prbm}
\end{figure}
Since our method closely follows the technical features of the DMQMC algorithm, the sign problem manifests itself in the same fashion, namely by a system specific population plateau. We performed ground-state simulations for the $4\times4$ triangular Heisenberg model which is known to be affected by the sign problem \cite{FCIQMC2}. Our simulations display the same qualitative behavior as the one reported in \cite{FCIQMC2} and \cite{DMQMC} using FCIQMC and DMQMC, respectively. At first an exponential growth of the triplet's population occurs due to a rapid spreading of the triplet over the Hilbert space. Then, because of competing contributions coming from triplets with opposite weight signs, the triplet's population stabilizes at a plateau height. Finally a second exponential growth stemming from a non-zero ground state energy emerges, signalling that the ground state has been reached and the shift update can be enabled. The energy and population evolution are plotted in Fig.\ \ref{fig:sgn_prbm}. The population plateau is the phase in the simulation during which the sign problem is overcome and it corresponds to the noisy section on the energy curve. As the plateau phase ends, the energy shift update is used to prevent the triplet's population unwanted growth. Data sampling begins at this point. The initiator approximation allows one to dramatically decrease the height of the population plateau by tuning the parameter $c_{\text{init}}$, without accumulating too large a systematic error. If the parameter $c_{\text{init}}$ is too large, the statistical errors will be smaller than the systematic one introduced by the initiator approximation itself. Generally speaking, the parameter $n_{\text{init}}$ is set to one, resulting in all the triplets with zero dynamic norm being initiators by default.
\begin{figure}
    \centering
    \includegraphics[width = \linewidth]{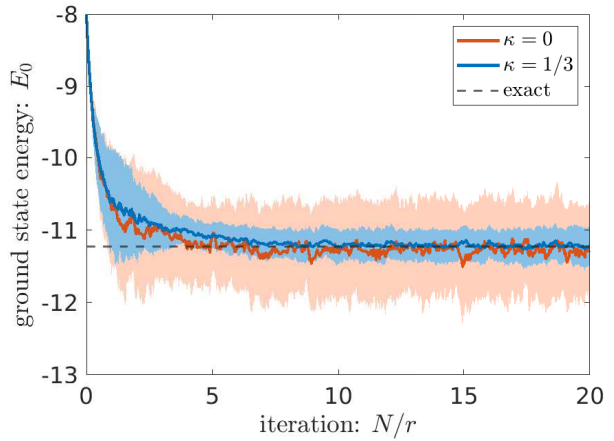}
    \caption{Ground state thermalization of the $4\times4$ square Heisenberg model. The population of both simulations is about $3.5\cdot 10^5$ psips, $r = 30$ and the initial population is $10^4$. The blue line was simulated with importance sampling whereas the red without.}
    \label{fig:importance_sampling}
\end{figure}
In order to test our importance sampling procedure further simulations of the thermalization phase for the $4\times 4$ square Heisenberg model have been performed. For this specific model calculating the dynamic norm is computationally expensive whenever it is is large. As it turns out, such a calculation is equivalent to a minimum weight perfect-matching problem which can be solved with a Blossom algorithm whose computational complexity scales as $\mathcal{O}(L^6)$ in the worst case scenario \cite{blossom}. In order to cut down the simulation time, we modify the definition of dynamic norm. For a triplet $(c,i,j)$, the new norm $n_{ij}$ measures the number of local spin exchanges between the state $i$ and state $j$. This can be easily implemented by a bit-wise operation and it is computationally very inexpensive (from and algorithmic point of  view, it is fully equivalent to a XOR operation).  The results of two independent simulations, one with importance sampling and one without, are depicted on Fig.\ \ref{fig:importance_sampling}. The final value of the triplet's population for both simulations is about $3.5 \times 10^5$. The statistical error on the value of the ground state energy after sampling is approximately $3$ times smaller in the simulation with importance sampling. Note that, if the spring constant $\kappa$ in Eq.~\eqref{eq:spring} is too large, some physically relevant triplets might be erroneously removed from the simulation. This, in turn, could lead to a severe underestimate of the triplet distribution and a failure to converge to the correct ground state.

\subsection{Case study: the Fermi-Hubbard model}
\label{Fermi}
The second model we use to benchmark our algorithm is the Fermi-Hubbard model on an $L\times L$ square lattice \cite{hubbard}. Similarly to the Heisenberg model, this is also of paramount importance as it is believed to describe several important phenomena in solid-state physics, {\textit e.g.} high-temperature superconductivity. Yet, a general exact solution is completely elusive and the Fermi-Hubbard model has been under investigation for several decades \cite{hubbardrev}.
The Fermi-Hubbard Hamiltonian reads
\begin{equation}
    H = -t\sum_{\sigma = \{\uparrow,\downarrow\}}\sum_{\langle a,b\rangle} {c_a^\sigma}^\dagger{c_b^\sigma} + U \sum_{a}n^\uparrow_an^\downarrow_a,
\end{equation}
where $a = 1, \dots, L^2$ are the lattice sites, $\langle a, b\rangle$ denotes nearest neighbours on the lattice, and $n_a^\sigma = {c_a^\sigma}^\dagger c_a^\sigma$ is the number operator for particles with spin $\sigma$ at site $a$. The fermionic ladder operators ${c_a^\sigma}^\dagger$, $c_a^\sigma$ follow the usual anti-commutation rules $\{{c_a^\sigma}^\dagger,c_b^{\sigma'}\} = \delta_{a,b}\delta_{\sigma, \sigma'}$. The total Hamiltonian is split into a free and an interaction part
\begin{equation}
    H^{\text{free}} = U \sum_{a}n^\uparrow_an^\downarrow_a, \qquad H^{\text{int}} = -t\sum_{\sigma = \{\uparrow,\downarrow\}}\sum_{\langle a,b\rangle} {c_a^\sigma}^\dagger{c_b^\sigma}.
\end{equation}

\begin{figure}
    \centering
    \includegraphics[width= \linewidth]{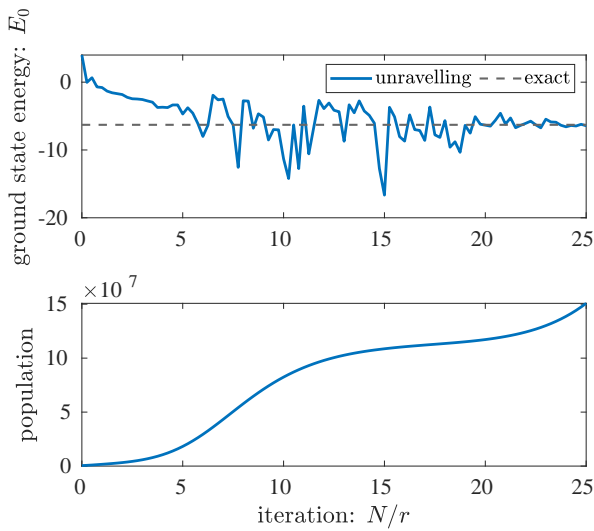}
    \caption{Ground state thermalization of the $3\times 3$ Hubbard model with 10 electrons, for $r=4$, $\kappa = 1/40$, $c_{\text{init}}=1$, $n_{\text{init}} = 1$.}
    \label{fig:hub3x3}
\end{figure}

An example of the thermalization stage of the $3\times 3$ Hubbard model for $U = 4$, $t = 1$ with 10 electrons is shown of Fig.\ \ref{fig:hub3x3}. The use of importance sampling, the initiator approximation and a large time-step allows to reduce the height of the plateau and the simulation time to a few minutes only. 

\begin{figure}
    \centering
    \includegraphics[width = \linewidth]{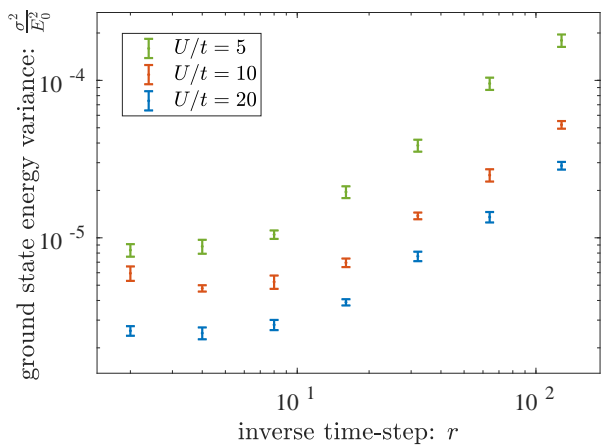}
    \caption{Relative variance $\sigma^2/E_0^2$ of the ground state energy $E_0$ as a function of the inverse time-step $r$ for the $1\times 10$ Hubbard model at half-filling with various interaction strengths. Each point consists of 10 independent simulations, each with the same number iteration $N=3.2\cdot 10^4$ and the same population $10^5$.}
    \label{fig:var}
\end{figure}
We show now that the error for a given simulation time decreases with $r$. As mentioned previously, no constraint on the lower bound of $r$ exists \textit{a priori}, since the rate of convergence depends on the initial condition. Indeed, in view of Eq.\ (\ref{eq:fund_disc}), if $\rho_0$ is chosen close to the fixed point $\rho_\infty$, convergence will only take a small number of iterations. It is therefore desirable to choose $\rho_0$ cleverly in order to decrease the number of iterations needed before the sampling phase. Furthermore, in our algorithm, a large imaginary-time step $1/r$ is more advantageous in order to reduce the statistical error on an observable's average. Since the sampling stage (the region $N/r >10$ on Fig.\ \ref{fig:importance_sampling}) produces a sequence of data points that are correlated to each other, we have to perform a block analysis to estimate the decorrelated variance $\sigma^2$ which, in turn, is related to the measurement's error bars. This correlated trajectory is characterized by two parameters, the decorrelation time $\tau$, related to the number of iterations $M$ between two decorrelated points via $M= \tau r$, and the amplitude of the fluctuations around the average, proportional to the jump amplitude $1/2r$. The large-scale fluctuations (that is, those at the lowest frequencies) appears to be independent of $r$ and, thus,
only an increase in the total triplet's population can reduce them. For a constant total number of sampled data points $\sigma^2 \sim A^2M\sim r$, with $A$ being the amplitude of the large-scale fluctuations. For a constant number of iterations and for an increasing time-step, the correlation time will become smaller, thus increasing the number of independent estimates. In Fig.\ \ref{fig:var} we show the variance of the ground state energy as a function of $r$. The linear unit slope is clearly visible on the right hand side of the figure while the transition to the minimum value is due to the decorrelation time approaching unity, when each data point is uncorrelated from the previous ones. Reducing $r$ will only increase the jump amplitude, resulting in an increasing error on the left hand side of the figure. This feature is particularly useful in a perturbative regime where $H^{\text{free}}\gg H^{\text{coll}}$ or $U \gg t$.

\section{Conclusions and open perspectives}
\label{Concs}

In this manuscript we have introduced a new quantum Monte Carlo method to investigate equilibrium properties of many-body systems. The method uses two-process unravellings to solve a piece-wise-deterministic stochastic process whose average reproduces the solution to the zero-temperature Bloch equation. Furthermore, it allows one to statistically sample the density matrix as an ensemble of triplets $\{(c_n,i_n,j_n)\}_n$ while restricting the Hilbert space exploration to physically relevant states only. Owning to the use of the Laplace transform and of a fixed-point iteration scheme, the unravelling algorithm is exact, excecpt for the initiator approximation. Thus, the rate at which the interaction Hamiltionian is applied to the ensemble of triplets has no \textit{a priori} lower bound. When chosen small enough, this can reduce the number of necessary iterations as well as the statistical error on the statistical averages. In general this work presents an interaction picture method, where the free part is integrated exactly due to the choice of the basis and where a small interaction allows a fast convergence. This comes from the fact that the initial state is close to the ground state, thus allowing to choose a very large time-step while still guaranteeing to reach the true ground state.
\par
In the light of the findings reported in this article, we are confident that our method can be applied to a number of different scenarios in quantum chemistry and condensed matter physics. On the other hand, we also foresee a series of future investigations. We believe the most pressing to be; i) the extension to real-time simulations, including out-of equilibrium dynamics; ii) the application of our method to the study of multi-correlation functions, crucial for understanding quantum correlations;  and finally iii) the inclusion of dissipation and decoherence to study thermalization properties. 
\par

\section*{Acknowledgements}
R.C. and H.C.O. would like to thank Elia Dietler for his ideas in numerous discussions and for providing valuable simulations.

\appendix

\section{Compression and decompression steps}
\label{sec:comp}
Prior to the execution of the loop the ensemble is modified as to improve the statistics without influencing directly the averages. This modification is carried out through \textit{compression} or \textit{decompression}. In a compression, classes of triplets are formed by grouping together all the triplets associated to a fixed pair of states, for instance, $(i,j)$. These are then replaced by a single triplet whose weight is equal to the sum of the weights of all the members of the class.
Decompression is applied on a compressed ensemble. 
A single class of triplets $(i,j)$ is split into triplets with unit-weight (in absolute value) $(\text{sgn}(c_n),i_n,j_n)$, and a single rest triplet $(c_r,i_n,j_n)$, with $c_r =\text{sgn}(c_n)(|c_n|-\lfloor |c_n|\rfloor)$ ($\lfloor \cdot \rfloor$ is the floor function). The rest triplet is then removed from the simulation with probability $1-|c_r|$; otherwise its weight is updated to $\text{sgn}(c_n)$. That way, the total statistical weight is conserved on average. For the case of initiators the following rules apply;
\paragraph*{Initiator decompression} A triplet $(c,i,j)$ whose dynamic norm is strictly lower that a critical value $n_{\textup{init}}$ or whose weight $c$ in absolute value is strictly larger than a critical weight $|c| > c_{\text{init}}$ is upgraded to initiator. After this step, standard decompression is performed.
\paragraph*{Initiator compression}
If a  $(i,j)$ class has a single representative that was spawned by a non-initiator within the same loop, it is  removed from the simulation. Otherwise all the triplets in the class $(i,j)$ are replaced by a single triplet representative whose weight is the sum of the other representative's weights. 

\section{Population control via shift update}
\label{sec:shiftupdate}
 Similarly to DMQMC and related methods, we control the triplet's population dynamics using the energy shift $S$ introduced earlier. Once the population has reached a desired steady level, the shift is updated according to the following rule 
\begin{equation}
    S_L = S_{L-1} - r\xi\log\left(P_L/P_{L-1}\right),
    \label{eq:s_update}
\end{equation}
where $S_L$ denotes the shift at loop $L$, $\xi$ is a damping parameter and $P_L$ is the population at loop $L$. During the simulation we keep track of the population right after the compression so that we know the number of spawning attempts that have been performed in the previous step. This update step stabilizes the population and guarantees that the energy shift $S(\beta)$ will converge to the ground state energy $E_0$. The closer the initial shift $S_0 > E_0$ is to the ground state energy, the slower the initial triplet's population will increase. In general, both $S(\beta)$ and $\xi$ will have the same effect on the population as in all related methods previously introduced in literature (see \cite{FCIQMC3} for more details).\\

\section{Flowchart of the algorithm}
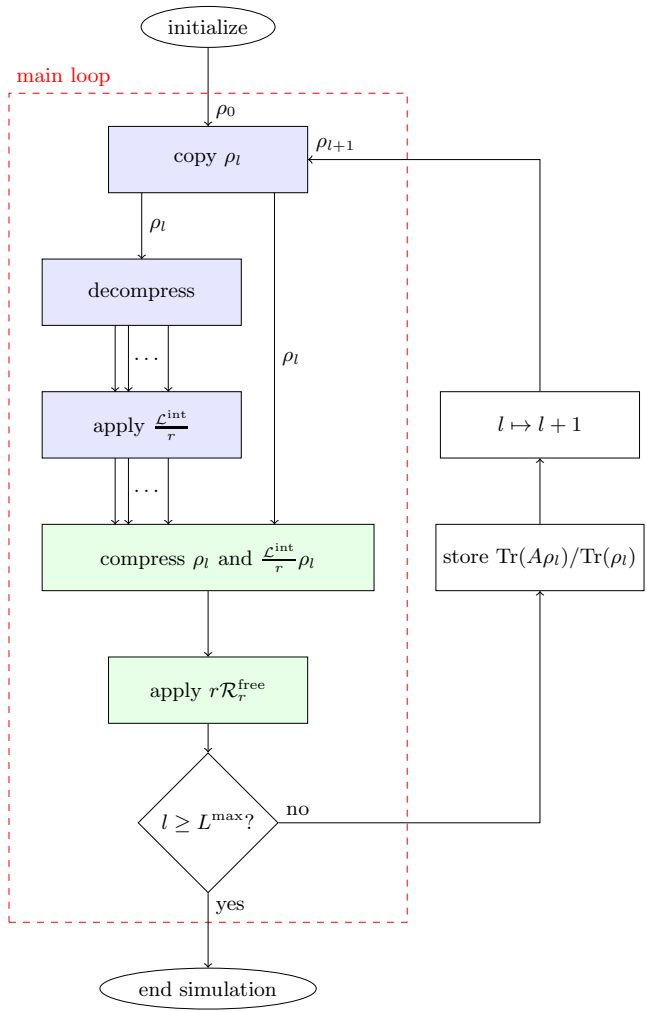
\begin{figure}[h!]
    \centering
    \resizebox{\linewidth}{!}{
        \begin{tikzpicture}
            \node[ellipse, draw] (init) at (0, 14) {initialize};
            \node[rectangle, draw, minimum width = 3cm, minimum height = 1cm, fill = blue!10!white] (copy) at (0, 12) {copy $\rho_l$};
            \node[rectangle, draw, minimum width = 3cm, minimum height = 1cm, fill = blue!10!white] (decomp) at (-1, 10) {decompress};
            \node[rectangle, draw, minimum width = 3cm, minimum height = 1cm, fill = blue!10!white] (int) at (-1,8) {apply $\frac{\mathcal{L}^{\text{int}}}{r}$};
            \node[rectangle, draw, minimum width = 5cm, minimum height = 1cm, fill = green!10!white] (comp) at (0, 6) {compress $\rho_l$ and $\frac{\mathcal{L}^{\text{int}}}{r}\rho_l$};
            \node[rectangle, draw, minimum width = 3cm, minimum height = 1cm, fill = green!10!white] (free) at (0,4) {apply $r\mathcal{R}_r^{\text{free}}$};
            \node[diamond, draw] (endloop) at (0,2) {$l \geq L^{\text{max}}$?};
            \node[ellipse, draw] (endsimul) at (0,-0.5) {end simulation};
            \node[rectangle, draw, minimum width = 3cm, minimum height = 1cm] (newloop) at (5,8) {$l \mapsto l+1$};
            \node[rectangle, draw, minimum width = 3cm, minimum height = 1cm] (measure) at (5,6) {store $\text{Tr}(A\rho_l)/\text{Tr}(\rho_l)$};
        
            \draw[dashed, red](-3,0.5) rectangle (3,13);
            \draw(-3,13) node[anchor= south west]{\textcolor{red}{main loop}};
        
            \draw[->] (init)--(copy);
            \draw[->] (comp)--(free);
            \draw[->] (free)--(endloop);
            \draw[->] (endloop)--(endsimul);
            \draw[->] (measure)--(newloop);
            \draw[->] (endloop)--(5, 2)--(measure);
            \draw(endloop.south) node[anchor=north west]{yes};
            \draw(endloop.east) node[anchor=south west]{no};
            \draw[->] (newloop)--(5, 12)--(copy);
            \draw(copy.east) node[anchor= south west]{$\rho_{l+1}$};
            \draw(copy.north) node[anchor= south west]{$\rho_{0}$};
            \draw[->] (1, 11.5)--node[midway, right]{$\rho_l$}(1, 6.5);
            \draw[->] (-1, 11.5)--node[midway, right]{$\rho_l$}(-1, 10.5);
            \draw[->] (-1.4, 9.5)--(-1.4,8.5);
            \draw[->] (-1.2, 9.5)--(-1.2,8.5);
            \draw[->] (-0.6, 9.5)--(-0.6,8.5);
            \draw(-0.9, 9)node{$\dots$};
        
            \draw[->] (-1.4, 7.5)--(-1.4,6.5);
            \draw[->] (-1.2, 7.5)--(-1.2,6.5);
            \draw[->] (-0.6, 7.5)--(-0.6,6.5);
            \draw(-0.9, 7)node{$\dots$};

        \end{tikzpicture}
    }
    \caption{Flowchart of the triplet unravelling algorithm. The main loop is in the red rectangle, the spawning step in blue and the free evolution in green. $\rho_l$ denotes the triplet ensemble at loop $l$. Multiple parallel lines (\textit{e.g.} between the decompression and interaction steps) illustrates the separation of triplets into child triplets.}
    \label{fig:flowchart}
\end{figure}

\bibliography{biblio.bib}

\end{document}